\documentclass[runningheads]{llncs}
\usepackage[hidelinks]{hyperref}
\usepackage{graphicx}
\usepackage{array}
\usepackage{amsmath}
\usepackage{xcolor}
\usepackage{makecell}
\usepackage{multirow}
\usepackage[misc]{ifsym}
\begin{document}
\title{MPBD-LSTM: A Predictive Model For Colorectal Liver Metastases Using Time Series Multi-phase Contrast-Enhanced CT Scans}
\titlerunning{MPBD-LSTM}

\author{Xueyang Li\inst{1} \and
Han Xiao\inst{2} \and
Weixiang Weng\inst{2}\and
Xiaowei Xu\inst{3}\and
Yiyu Shi\inst{1}$^{\textrm{(\Letter)}}$}

\authorrunning{X. Li et al.}

\institute{University of Notre Dame,
Notre Dame, USA \\ \email{\{xli34,yshi4\}@nd.edu} \and
The First Affiliated Hospital of Sun Yat-sen University, Guangzhou, China\\
\email{xiaoh69@mail.sysu.edu.cn, wengwx3@mail2.sysu.edu.cn} \and
Guangdong Provincial People's Hospital, Guangzhou, China \\
\email{xiao.wei.xu@foxmail.com}}
\maketitle              
\begin{abstract}
Colorectal cancer is a prevalent form of cancer, and many patients develop colorectal cancer liver metastasis (CRLM) as a result. Early detection of CRLM is critical for improving survival rates. Radiologists usually rely on a series of multi-phase contrast-enhanced computed tomography (CECT) scans done during follow-up visits to perform early detection of the potential CRLM. These scans form unique five-dimensional data (time, phase, and axial, sagittal, and coronal planes in 3D CT). Most of the existing deep learning models can readily handle four-dimensional data (e.g., time-series 3D CT images)
and it is not clear how well they can be 
extended to handle the additional dimension of phase. 
In this paper, we build a dataset of time-series CECT scans 
to aid in the early diagnosis of CRLM, and build upon state-of-the-art deep learning 
techniques to evaluate how to best predict CRLM. Our 
experimental results show that a multi-plane architecture 
based on 3D bi-directioal LSTM, which we call MPBD-LSTM, works best, achieving an area under curve (AUC) of 0.79. On the other hand, 
analysis of the results shows that 
there is still great room  for further improvement. Our code is available at \url{https://github.com/XueyangLiOSU/MPBD-LSTM}.



\end{abstract}
\section{Introduction}
Colorectal cancer is the third most common malignant tumor, and nearly half of all patients with colorectal cancer develop liver metastasis during the course of the disease~\cite{hao2022predicting,yu2020emerging}. Liver metastases after surgery of colorectal cancer is the major cause of disease-related death. Colorectal cancer liver metastases (CRLM) have therefore become one of the major focuses in the medical field. Patients with colorectal cancer typically undergo contrast-enhanced computed tomography (CECT) scans multiple times during follow-up visits after surgery for early detection of CRLM, generating a 5D dataset. In addition to the axial, sagittal, and coronal planes in 3D CT scans, the data comprises contrast-enhanced multiple phases as its 4th dimension, along with different timestamps as its 5th dimension. Radiologists heavily rely on this data to detect the CRLM in the very early stage~\cite{xu2011imaging}. 




Extensive existing works have demonstrated the power of deep learning 
on various spatial-temporal data, and can potentially be applied 
towards the problem of CRLM. For example, originally 
designed for natural data, several mainstream models such as E3D-LSTM~\cite{wang2019eidetic}, ConvLSTM~\cite{shi2015convolutional} and PredRNN~\cite{wang2017predrnn} use Convolutional Neural Networks (CNN) to capture spatial features and Long Short-Term Memory (LSTM) to process temporal features. 
Some other models, such as SimVP~\cite{gao2022simvp}, replace LSTMs with CNNs but still have the capability of processing spatiotemporal information. 
These models can be adapted for classification tasks with the use of 
proper classification head. 

However, all these methods have only demonstrated 
their effectiveness towards 3D/4D data (i.e., time-series 2D/3D images), and 
it is not clear how to best extend them to work with the 5D CECT data. 
Part of the reason is due to the lack of public availability of such data. When extending these models towards 5D CECT data, some decisions need to be made, for example: 1) What is the most effective way to incorporate the phase information? Simply concatenating different phases together may not be the optimal choice, because the positional information of the same CT slice in different phases would be lost. 2) Shall we use uni-directional LSTM or bi-direction LSTM? E3D-LSTM~\cite{wang2019eidetic} shows uni-directional LSTM works well on natural videos while several other works show bi-directional LSTM is needed in certain medical image segmentation tasks ~\cite{kim2021bidirectional,chen2016combining}. 




In this paper, we investigate how state-of-art deep learning models can be applied to the CRLM prediction task using our 5D CECT dataset. We evaluate the effectiveness of bi-directional LSTM and explore the possible method of incorporating different phases in the CECT dataset. Specifically, we show that the best prediction accuracy can be achieved by enhancing E3D-LSTM~\cite{wang2019eidetic} with a bi-directional LSTM and a multi-plane structure. 


\section{Dataset and Methodology}
\subsection{Dataset}

\begin{figure}
\includegraphics[width=\textwidth]{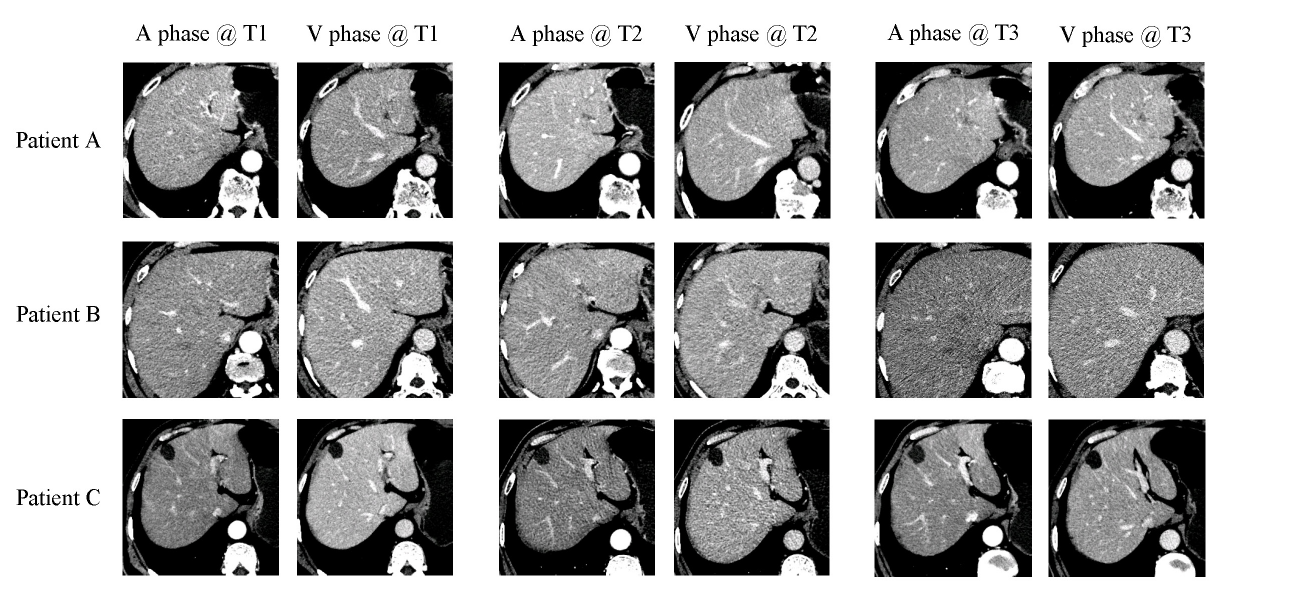}
\caption{Representative slices from 3D CT images of different patients in our dataset, at A/V phases and timestamps T0, T1, T2 (cropped to 256x256 for better view).} \label{vapic}
\end{figure}

\setlength{\tabcolsep}{0.5em}
\begin{table}
\centering
\caption{Characreristics of our dataset}\label{tab0}
\begin{tabular}{ccccc}
\hline
Cohort & \# of positive cases & \# of negative cases & total cases & positive rate  \\
\Xhline{2\arrayrulewidth}
1st & 60 & 141 & 201 & 0.299 \\
\hline
2nd & 9 & 59 & 68 & 0.132 \\
\hline
{\bfseries Total} & {\bfseries 69} & {\bfseries 200} & {\bfseries 269}  & {\bfseries 0.257} \\
\hline
\end{tabular}
\end{table}

When patients undergo CECT scans to detect CRLM, typically three phases are captured: the unenhanced plain scan phase (P), the portal venous phase (V), and the arterial phase (A). The P phase provides the basic shape of the liver tissue, while the V and A phases provide additional information on the liver's normal and abnormal blood vessel patterns, respectively~\cite{patel2022ct}. Professional radiologists often combine the A and V phases to determine the existence of metastases since blood in the liver is supplied by both portal venous and arterial routes.

Our dataset follows specific inclusion criteria:
\begin{itemize}
    \item \textbf{No tumor} appears on the CT scans. That means patients \textbf{have not been diagnosed} as CRLM when they took the scans.
    \item Patients were previously diagnosed with colorectal cancer TNM stage I to stage III, and recovered from colorectal radical surgery.
    \item Patients have two or more times of CECT scans.
    \item We already determined whether or not the patients had liver metastases within 2 years after the surgery, and manually labeled the dataset based on this.
    \item No potential focal infection in the liver before the colorectal radical surgery. 
    \item No metastases in other organs before the liver metastases.
    \item No other malignant tumors.
\end{itemize}
Our retrospective dataset includes two cohorts from two hospitals. The first cohort consists of 201 patients and the second cohort includes 68 patients. Each scan contains three phases and 100 to 200 CT slices with a resolution of $512 \times 512$. Patients may have different numbers of CT scans, ranging from 2 to 6, depending on the number of follow-up visits. CT images are collected with the following acquisition parameters: window width 150, window level 50, radiation dose 120 kV, slice thickness 1 mm, and slice gap 0.8 mm. All images underwent manual quality control to exclude any scans with noticeable artifacts or blurriness and to verify the completeness of all slices. Additional statistics on our dataset are presented in Table~\ref{tab0} and examples of representative images are shown in Fig.~\ref{vapic}. The dataset is available upon request.

\begin{figure}
\includegraphics[width=\textwidth]{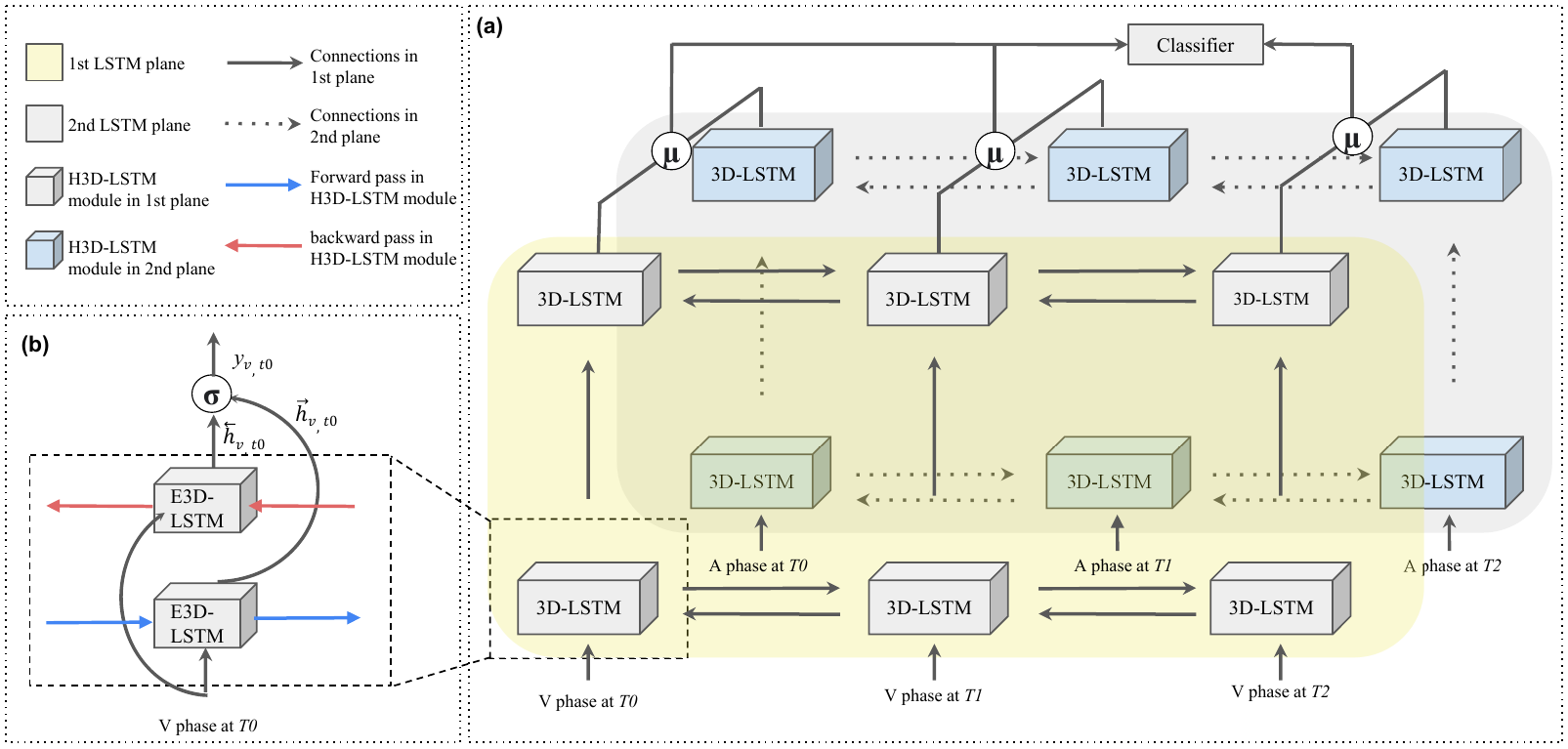}
\caption{(a) The general structure of MPBD-LSTM. The yellow plane is the 1st plane which is used to process the portal venous phase CT scans, and the gray plane is the second one used to process the arterial phase CT scans. $\mu$ is the average function. (b) The inner structure of a 3D-LSTM module. \textcolor{blue}{Blue arrow} stands for the forward pass which generates the output of $\overrightarrow{h}_{v,t_0}$ and \textcolor{red}{red arrow} indicates the backward pass generating the output of $\overleftarrow{h}_{v,t_0}$. $\sigma$ is the function used to combine two hidden-state outputs. $y_{v,t_0}$ is the output of this 3D-LSTM module after processed by $\sigma$.} \label{model}
\end{figure}

\subsection{Methods}
Numerous state-of-the-art deep learning models are available to effectively process 4D data. In this paper, we will evaluate some of the most popular ones: 

1) SaConvLSTM, introduced by Lin \textit{et al.}~\cite{lin2020self}, incorporates the self-attention mechanism into the ConvLSTM\cite{shi2015convolutional} structure, which improves the ability to capture spatiotemporal correlations compared to traditional LSTM. 

2) E3D-LSTM, introduced by Wang \textit{et al.}~\cite{wang2019eidetic}, integrates 3D CNNs into LSTM cells to capture both short- and long-term temporal relations. They used 3D-CNNs to handle the 3D data at each timestamp and LSTMs to compute information at different timestamps. 

3)PredRNN-V2, introduced by Wang \textit{et al.}~\cite{wang2017predrnn,wang2021predrnn}, uses Spatiotemporal LSTM (ST-LSTM) by stacking multiple ConvLSTM units and connecting them in a zigzag pattern to handle spatiotemporal data of 4 dimensions.

4) SimVP~\cite{gao2022simvp}, introduced by \textit{Gao et al.}, uses CNN as the translator instead of LSTM. 

All of these models need to be modified to handle 5D CECT datasets. A straightforward way to extend them is simply concatenating the A phase and V phase together, thus collapsing the 5D dataset to 4D. However, such an extension 
may not be the best way to incorporate the 5D spatiotemporal information, because the positional information of the same CT slice in different phases would be lost. 
Below we explore an alternative modification \textit{multi-plane bi-directional LSTM} (MPBD-LSTM), based on E3D-LSTM,  
to handle the 5D data. 

\subsubsection{MPBD-LSTM}
The most basic building block in MPBD-LSTM is the 3D-LSTM modules. Each 3D-LSTM module is composed of two E3D-LSTM cells~\cite{wang2019eidetic}. Additionally, inspired by the bi-directional LSTM used in medical image segmentation task~\cite{chen2016combining}, we replace the uni-directional connections with bi-directional connections by using the backward pass in the 2nd E3D-LSTM cell in each 3D-LSTM module. This allows us to further jointly compute information from different timestamps and gives us more accurate modeling of temporal dynamics. The inner structure of one such module is shown in Fig.~\ref{model}(b). Aside from the two E3D-LSTM cells, it also includes an output gate $\sigma$. Each 3D-LSTM module will generate an output $y_{v,t}$, which can be calculated as~\cite{cui2018deep}:
\begin{equation}
y_{v,t} = \sigma(\overrightarrow{h}_{v,t}, \overleftarrow{h}_{v,t})
\end{equation}
where $\overrightarrow{h}_{v,t}$ and $\overleftarrow{h}_{v,t}$ are the output hidden state of the forward pass and backward pass of phase $v$ at timestamp $t$, and $\sigma$ is the function which is used to combine these two outputs, which we choose to use a summation function to get the summation product of these two hidden states. Therefore, the output of the bi-directional LSTM module presented in Fig.~\ref{model}(b) can be represented as:
\begin{equation}
y_{v,t_0} = \overrightarrow{h}_{v,t_0} \oplus \overleftarrow{h}_{v,t_0}
\end{equation}
in which $\oplus$ stands for summation. After this, the output $y_{v,t_0}$ is passed into the bi-directional LSTM module in the next layer and viewed as input for this module.

Fig.~\ref{model}(a) illustrates how MPBD-LSTM uses these 3D-LSTM building blocks to handle the multiple phases in our CT scan dataset.  We use two planes, one for the A phase and one for the V phase, each of which is based on a backbone of E3D-LSTM~\cite{wang2019eidetic} with the same hyperparameters.
We first use three 3D-CNN encoders (not displayed in Fig.~\ref{model}(a)) as introduced in E3D-LSTM to extract the features. Each encoder is followed by a 3D-LSTM stack (the ``columns'') that processes the spatiotemporal data for each timestamp. The stacks are bidirectionally connected, as we described earlier, and consist of two layers of 3D-LSTM modules that are connected by their hidden states. When the spatiotemporal dataset enters the model, it is divided into smaller groups based on timestamps and phases. The 3D-LSTM stacks process these groups in parallel, ensuring that the CT slices from different phases are processed independently and in order, preserving the positional information. After the computation of the 3D-LSTM modules in each plane, we use an average function to combine the output hidden states from both planes.

An alternative approach is to additionally connect two planes by combining the hidden states of 3D-LSTM modules and taking their average if a module receives two inputs. However, we found that such design actually resulted in a worse performance. This issue will be demonstrated and discussed later in the ablation study.

In summary, the MPBD-LSTM model comprises two planes, each of which contains three 3D-LSTM stacks with two modules in each stack. It modifies E3D-LSTM by using bi-directional connected LSTMs to enhance communication between different timestamps, and a multi-plane structure to simultaneously process multiple phases.

\section{Experiments}
\subsection{Data augmentation and selection} We selected 170 patients who underwent three or more CECT scans from our original dataset, and cropped the images to only include the liver area, as shown in Fig.~\ref{vapic}. Among these cases, we identified 49 positive cases and 121 negative cases. 
To handle the imbalanced training dataset, we selected and duplicated 60\% of positive cases and 20\% of negative cases by applying Standard Scale Jittering (SSJ)\cite{ghiasi2021simple}. 
For data augmentation, we randomly rotated the images from -30 degrees to 30 degrees and employed mixup \cite{zhang2017mixup}. We applied the same augmentation technique consistently to all phases and timestamps of each patient's data. We also used Spline Interpolated Zoom (SIZ)~\cite{zunair2020uniformizing} to uniformly select 64 slices. For each slice, the dimension was $256 \times 256$ after cropping. 
We used the A and V phases of CECT for our CRLM prediction task since the P phase is only relevant when tumors are significantly present, which is not the case in our dataset. 
The dimension of our final input is $(3\times2\times64\times64\times64)$, representing $(T\times P\times D\times H\times W)$, where $T$ is the number of timestamps, $P$ is the number of different phases, $D$ is the slice depth, $H$ is the height, and $W$ is the width.

\subsection{Experiment setup} As the data size is limited, 10-fold cross-validation is adopted, and the ratio of training and testing dataset is 0.9 and 0.1, respectively. Adam optimizer~\cite{kingma2014adam} and Binary Cross Entropy loss function~\cite{ba2014deep} are used for network training. For MPBD-LSTM, due to GPU memory constraints, we set the batch size to one and the number of hidden units in LSTM cells to 16, and trained the model till converge with a learning rate of 5e-4. Each training process required approximately 23 GB of memory and took about 20 hours on an Nvidia Titan RTX GPU. We ran the 10 folds in parallel on five separate GPUs, which allowed us to complete the entire training process in approximately 40 hours. We also evaluated E3D-LSTM~\cite{wang2019eidetic}, PredRNN-V2~\cite{wang2021predrnn}, SaConvLSTM~\cite{lin2020self}, and SimVP~\cite{gao2022simvp}. As this is a classification task, we evaluate all models' performance by their AUC scores.

\section{Results and Discussion}

\begin{table}
\centering
\caption{AUC scores of different models on our dataset}\label{result}
\begin{tabular}{>{\centering\arraybackslash}p{7cm} >{\centering\arraybackslash}p{3cm}}
\hline
{\bfseries  Model}&  {\bfseries  AUC score } \\
\hline
E3D-LSTM~\cite{wang2019eidetic} &  0.755  \\
SaConvLSTM~\cite{lin2020self} & 0.721 \\
PredRNN-V2~\cite{wang2021predrnn} & 0.765 \\
SimVP~\cite{gao2022simvp} & 0.662 \\
MPBD-LSTM& {\bfseries 0.790}\\
\hline
\end{tabular}
\end{table}


 \noindent Table~\ref{result} shows the AUC scores of all models tested on our dataset. Additional data on accuracy, sensitivity specificity, etc can be found in the supplementary material. The MPBD-LSTM model outperforms all other models with an AUC score of 0.790. Notably, SimVP~\cite{gao2022simvp} is the only CNN-based model we tested, while all other models are LSTM-based. Our results suggest that LSTM networks are more effective in handling temporal features for our problem compared with CNN-based models. Furthermore, PredRNN-V2~\cite{wang2021predrnn}, which passes memory flow in a zigzag manner of bi-directional hierarchies, outperforms the uni-directional LSTM-based SaConvLSTM~\cite{lin2020self}. Although the architecture of PredRNN-V2 is different from MPBD-LSTM, it potentially supports the efficacy of jointly computing spatiotemporal relations in different timestamps.

\begin{table}
\centering
\caption{Ablation study on bi-directional connection and multi-planes}\label{Ablation}
\begin{tabular}{>{\centering\arraybackslash}p{7cm} >{\centering\arraybackslash}p{4cm}}
\hline
{\bfseries  Model}&  {\bfseries  AUC score } \\
\hline
MPBD-LSTM w/o multi-plane  & 0.774 \\
MPBD-LSTM w/o bi-directional connection & 0.768 \\
MPBD-LSTM w/ inter-plane connections & 0.786\\
MPBD-LSTM& {\bfseries 0.790}\\
\hline
\end{tabular}
\end{table}

\subsubsection{Ablation study on model structures}
As shown in Table~\ref{Ablation}, to evaluate the effectiveness of multi-plane and bi-directional connections, we performed ablation studies on both structures. First, we removed the multi-plane structure and concatenated the A and V phases as input. This produced a one-dimensional bi-directional LSTM (Fig.~\ref{model}(a), without the gray plane) with an input dimension of $3\times128\times64\times64$, which is the same as we used on other models. The resulting AUC score of 0.774 is lower than the original model's score of 0.790, indicating that computing two phases in parallel is more effective than simply concatenating them. After this, we performed an ablation study to assess the effectiveness of the bi-directional connection. By replacing the bi-directional connection with a uni-directional connection, the MPBD-LSTM model's performance decreased to 0.768 on the original dataset. This result indicates that the bi-directional connection is crucial for computing temporal information effectively, and its inclusion is essential for achieving high performance in MPBD-LSTM.

Also, as mentioned previously, we initially connected the 3D-LSTM modules in two planes with their hidden states. However, as shown in Table~\ref{Ablation}, we observed that inter-plane connections actually decreased our AUC score to 0.786 compared to 0.790 without the connections. This may be due to the fact that when taking CT scans with contrast, different phases have a distinct focus, showing different blood vessels as seen in Fig.~\ref{vapic}. Connecting them with hidden states in the early layers could disrupt feature extraction for the current phase. Therefore, we removed the inter-plane connections in the early stage, since their hidden states are still added together and averaged after they are processed by the LSTM layers.

\begin{table}
\centering
\caption{Ablation study on timestamps and phases }\label{Ablationtimes}
\begin{tabular}{>{\centering\arraybackslash}p{7cm} >{\centering\arraybackslash}p{4cm}}
\hline
{\bfseries  Model structure}&  {\bfseries  AUC score } \\
\hline
MPBD-LSTM @ T0 & 0.660 \\
MPBD-LSTM @ T1 & 0.676 \\
MPBD-LSTM @ T2 & 0.709\\
MPBD-LSTM @ all timestamps w/ only A phase & 0.653\\
MPBD-LSTM @ all timestamps w/ only V phase & 0.752\\
MPBD-LSTM @ All 3 timestamps& {\bfseries 0.790}\\
\hline
\end{tabular}
\end{table}

\subsubsection{Ablation study on timestamps and phases}
We conducted ablation studies using CT images from different timestamps and phases to evaluate the effectiveness of time-series data and multi-phase data. The results, as shown in Table~\ref{Ablationtimes}, indicate that MPBD-LSTM achieves AUC scores of 0.660, 0.676, and 0.709 if only images from timestamps T0, T1, and T2 are used, respectively. These scores suggest that predicting CRLM at earlier stages is more challenging since the features about potential metastases in CT images get more significant over time. However, all of these scores are significantly lower than the result using CT images from all timestamps. This confirms the effectiveness of using a time-series predictive model. Additionally, MPBD-LSTM obtains AUC scores of 0.653 and 0.752 on single A and V phases, respectively. These results suggest that the V phase is more effective when predicting CRLM, which is consistent with medical knowledge~\cite{xu2011imaging}. However, both of these scores are lower than the result of combining two phases, indicating that a multi-phase approach is more useful.

\subsubsection{Error Analysis}
In Fig.~\ref{vapic}, Patients B and C are diagnosed with positive CRLM later. MPBD-LSTM correctly yields a positive prediction for Patient B with a confidence of 0.82, but incorrectly yields a negative prediction for Patient C with a confidence of 0.77. With similar confidence in the two cases, the error is likely due to the relatively smaller liver size of Patient C. 
Beyond this case, we find that small liver size is also present in most of the false negative cases. A possible explanation would be that smaller liver may provide less information for accurate prediction of CRLM. How to effectively address inter-patient variability in the dataset, perhaps by better fusing the 5D features, requires further research from the community 
in the future.



\section{Conclusion}
In this paper, we put forward a 5D CECT dataset for CRLM prediction. Based on the popular E3D-LSTM model, we established MPBD-LSTM model by replacing the uni-directional connection with the bi-directional connection to better capture the temporal information in the CECT dataset. Moreover, we used a multi-plane structure to incorporate the additional phase dimension. MPBD-LSTM achieves the highest AUC score of 0.790 among state-of-the-art approaches. Further research is still needed to improve the AUC.

\bibliographystyle{splncs04}
\bibliography{bibliography}

\end{document}